\documentclass[runningheads]{llncs}
\usepackage[utf8]{inputenc}

\usepackage{hyperref}
\usepackage{longtable}
\usepackage{subcaption}
\usepackage{graphicx}

\usepackage{listings}
\usepackage[dvipsnames]{xcolor}
\usepackage{textcomp}
\usepackage{lstcoq}
\usepackage{supertabular}
\usepackage{amsmath,amssymb}
\usepackage{wrapfig}

\newcommand{\lbracett}{\texttt{\{}}
\newcommand{\rbracett}{\texttt{\}}}

\lstdefinelanguage{michelson}{%
   columns=fullflexible,%
   basicstyle=\small\tt ,
   commentstyle=\slshape,%
   keywords={%
     \{,\},
     DROP, DUP, SWAP, PUSH, SOME, NONE, UNIT, IF_NONE,%
   PAIR, CAR, CDR, LEFT, RIGHT, IF_LEFT, IF_RIGHT, NIL,%
   CONS, IF_CONS, SIZE, EMPTY_SET, EMPTY_MAP, MAP, ITER,%
   MEM, GET, UPDATE, IF, LOOP, LOOP_LEFT, LAMBDA, EXEC,%
   DIP, FAILWITH, CAST, RENAME, CONCAT, SLICE, PACK,%
   UNPACK, ADD, SUB, MUL, EDIV, ABS, NEG, LSL, LSR,%
   OR, AND, XOR, NOT, COMPARE, EQ, NEQ, LT, GT, LE,%
   GE, SELF, CONTRACT, TRANSFER_TOKENS, SET_DELEGATE,%
   CREATE_ACCOUNT, CREATE_CONTRACT, CREATE_CONTRACT,%
   IMPLICIT_ACCOUNT, NOW, AMOUNT, BALANCE, CHECK_SIGNATURE,%
   BLAKE, SHA, SHA, HASH_KEY, STEPS_TO_QUOTA, SOURCE,%
   SENDER, ADDRESS,%
   CMPEQ,CMPNEQ,CMPLT,CMPGT,CMPLE,CMPGE,%
   IFEQ,IFNEQ,IFLT,IFGT,IFLE,IFGE,%
   IFCMPEQ,IFCMPNEQ,IFCMPLT,IFCMPGT,IFCMPLE,IFCMPGE,%
   FAIL,%
   ASSERT,%
   ASSERT_EQ,ASSERT_NEQ,ASSERT_LT,ASSERT_LE,ASSERT_GT,ASSERT_GE,%
   ASSERT_CMPEQ,ASSERT_CMPNEQ,ASSERT_CMPLT,ASSERT_CMPLE,ASSERT_CMPGT,ASSERT_CMPGE,%
   ASSERT_NONE,ASSERT_SOME,%
   ASSERT_LEFT,ASSERT_RIGHT,%
   UNPAIR,%
   },%
   alsoletter={'},
   upquote=true,
   keywordstyle={\bfseries\sffamily},%
   morekeywords=[2]{%
     key, unit, signature, option, list, set, operation, address,%
     contract, pair, or, lambda, big_map, map,%
     int, nat, string, bytes, mutez, bool, key_hash, %
     timestamp, 'a, 'b, 'S, 'p%
   },%
   keywordstyle=[2]{\bfseries\ttfamily},%
   classoffset=2,%
   morekeywords=[3]{%
     storage, parameter, code %
   },%
   keywordstyle=[3]{\bfseries},%
   sensitive,%
   comment=[l]\#,%
   morestring=[d]",
   literate={->}{{$\rightarrow{}$}}1%
}[keywords,comments,strings]%

\lstMakeShortInline[language=michelson,basicstyle=\ttfamily,keywordstyle=\bfseries\sffamily\small]!

\lstdefinelanguage{albert}{%
   columns=fullflexible,%
   basicstyle=\tt ,
   commentstyle=\slshape,%
   keywordstyle={\color{red}\sffamily},%
   keywords={%
     \{,\},
     type, def,
     dup, drop,
     car, cdr,
     match, with, end,
     assert_some, Some, None, True, False, Left, Right,
     for, map, loop_left, in, do, done,
     failwith,
     contract, address, implicit\_account, amount,
     noop
   },%
   alsoletter={'},
   keywordstyle={\color{purple}\sffamily},%
   morekeywords=[2]{%
     key, unit, signature, option, list, set, operation, address,%
     contract, pair, or, lambda, big_map, map,%
     int, nat, string, bytes, mutez, bool, key_hash, %
     timestamp%
   },%
   keywordstyle=[2]{\color{blue}\ttfamily},%
   classoffset=2,%
   morekeywords=[3]{%
     storage, parameter, code %
   },%
   keywordstyle=[3]{\bfseries},%
   sensitive,%
   comment=[l]\#,%
   morestring=[d]",
   literate={->}{{$\rightarrow$}}1%
}[keywords,comments,strings]%

\lstMakeShortInline[language=albert]?

\begin{document}
\title{Albert, an intermediate smart-contract language for the Tezos blockchain}
%
%
\author{Bruno Bernardo \and
Rapha\"el Cauderlier \and
Basile Pesin \and
Julien Tesson}
\authorrunning{B. Bernardo \and R. Cauderlier \and B. Pesin \and J. Tesson}
%
\institute{Nomadic Labs, Paris, France\\ \email{\{first\_name.last\_name\}@nomadic-labs.com}}
\maketitle              
%
\begin{abstract}
  Tezos is a smart-contract blockchain. Tezos smart contracts are
  written in a low-level stack-based language called Michelson. In
  this article we present Albert, an intermediate language for Tezos
  smart contracts which abstracts Michelson stacks as linearly typed
  records. We also describe its compiler to Michelson, written in Coq,
  that targets Mi-Cho-Coq, a formal specification of Michelson
  implemented in Coq. 
\end{abstract}

\section{Introduction}
Tezos is an account-based public blockchain and smart-contract
platform. It was launched in June 2018 and an open-source
implementation is available~\cite{tezosGitLab}. The Tezos blockchain
distinguishes itself through its on-chain amendment procedure by which a
super-majority of stakeholders can modify a large part of the codebase, through
its liquid Proof-of-Stake consensus
algorithm~\cite{tezosLiquidPos}, and through its focus on formal methods
which is especially visible in the design and implementation of
Michelson, its smart-contract language.

Indeed, the Michelson interpreter is implemented using a GADT that statically
ensures the subject reduction property.  Moreover, Michelson is
formally specified in the Coq proof assistant. This Coq specification
is called Mi-Cho-Coq~\cite{michocoqgitlab} and its main application
today is the certification of Michelson smart contracts by deductive
verification~\cite{michocoq}.

However, the stack paradigm used by Michelson is too
low-level for complex applications. For this reason, several
high-level languages have been
developed~\cite{ligo,smartpy,fi,archetype,scaml,juvix}. Unfortunately,
their compilers to Michelson are not formally verified which limits
the application of formal methods for these languages.

In this article, we propose an intermediate language named Albert to
avoid the duplication of effort put into compilers to Michelson and to
ease the certification of these compilers.  The main feature of Albert
is that the Michelson stack is abstracted through named variables. The
duplication and destruction of resources are however explicit
operations in both Albert and Michelson, this is reflected in Albert
by the use of a linear type system.

We have formally specified the Albert language in the Ott
tool~\cite{ottLang} from which the Albert lexer, parser, and \LaTeX{}
documentation are generated. Ott can also generate typing and semantic
rules for Coq and other proof assistants. We have written the Albert
compiler in Coq as a function from the generated Coq output for the
Albert grammar to the Michelson syntax defined in Mi-Cho-Coq.

This article is organised as follows: Section~\ref{sec:michelson}
gives an overview of the Michelson smart-contract
language. Section~\ref{sec:albert} presents the Albert intermediate
language, the figures of this section have been produced by the
\LaTeX{} output of Ott. The Albert compiler is then presented in
Section~\ref{sec:compiler}. Section~\ref{sec:related-work} discusses
some related work and finally Section~\ref{sec:limits-future-work}
concludes the article by listing directions for future work.

The Albert specification and compiler are available at
\url{https://gitlab.com/nomadic-labs/albert/tree/WTSC20}.


\section{Overview of Michelson}
\label{sec:michelson}
Smart contracts are Tezos accounts of a particular kind.  They have
private access to a memory space on the chain called the
\emph{storage} of the smart contract, each transaction to a smart
contract account contains some data, the \emph{parameter} of the
transaction, and a \emph{script} is run at each transaction to decide
if the transaction is valid, update the smart contract storage, and
possibly emit new operations on the Tezos blockchain.

Michelson is the language in which the smart contract scripts are
written. The most important parts of the implementation of Michelson,
the typechecker and the interpreter, belong to the economic ruleset of
Tezos which evolves through the Tezos on-chain amendment voting
process.

\subsection{Design rationale}
\label{sec:design-rational}

Smart contracts operate in a very constrained context: they need to be
expressive, evaluated efficiently, and their resource consumption should be
accurately measured in order to stop the execution of programs that
would be too greedy, as their execution time impacts the block
construction and propagation.
Smart contracts are non-updatable programs that can handle valuable
assets, there is thus a need for strong guarantees on the correctness
of these programs.

The need for efficiency and more importantly for accurate account of
resource consumption leans toward a low-level interpreted language, while the
need for contract correctness leans toward a high-level, easily
auditable, easily formalisable language, with strong static
guarantees.

To satisfy these constraints, Michelson was made a Turing-complete,
low-level, stack based interpreted language (\textit{\`a la} Forth), facilitating the
measurement of computation costs, but with some high-level features \textit{\`a la} ML:
polymorphic products, options, sums, lists, sets and maps data-structures with
collection iterators, cryptographic primitives and anonymous functions.
Contracts are pure functions that take a stack as input and return a stack as output. This side-effect free design is an asset
for the conception of verification tools.

The language is statically typed to ensure the well-formedness
of the stack at any point of the program. This means that if a
program is well typed, and if it is being given a well-typed stack that matches its
 input expectation, then at any point of the program execution, the
given instruction can be evaluated on the current stack.

Moreover, to ease the formalisation of Michelson, ambiguous or hidden
behaviours have been avoided. In particular, unbounded integers are
used to avoid arithmetic overflows and division returns an option
(which is !None! if and only if the divisor is 0) so that the
Michelson programmer has to specify the behaviour of the program in
case of division by 0; she can however still \emph{explicitly} reject
the transaction using the !FAILWITH! Michelson instruction.

\subsection{Quick tour of the language}
\label{sec:quick-tour-language}

The full language syntax, type system, and
semantics are documented in~\cite{michelsonwhitedoc}, we give
here a quick and partial overview of the language.

\subsubsection{Contracts' shape}
\label{sec:contracts-shape}

A Michelson smart contract script is written in three parts: the
parameter type, the storage type, and the code of the contract.  A
contract's code consists of one block that can only be called with one
parameter, but multiple entry points can be encoded by branching on a
nesting of sum types and multiple parameters can be paired into one.

When the contract is deployed (or \emph{originated} in Tezos lingo) on the chain, it is bundled
with a data storage which can then only be changed by a contract's successful
execution.
The parameter and the storage associated to the contract are paired
and passed to the contract's code at each execution.
The execution of the code must return
a list of operations and the updated storage.

Seen from the outside, the type of the contract is the type of its
parameter, as it is the only way to interact with it.

\subsubsection{Michelson Instructions}
\label{sec:instruction-type}

As usual in stack-based languages, Michelson instructions take their
parameters on the stack. All Michelson instructions are typed as a
function going from the expected state of the stack, before the
instruction evaluation, to the resulting stack. For example, the
!AMOUNT! instruction used to obtain the amount in $\mu tez$
(\textit{i.e.} a millionth of a \textit{tez}, the smallest token unit
in Tezos) of the current transaction has type !'S -> mutez:'S! meaning
that for any stack type !'S!, it produces a stack of type !mutez:'S!.
Michelson uses an ordered type system which means that the number of
times values are used and the order in which they are introduced and
consumed matter and are visible at the type level. Some operations
such as !SWAP :: 'a:'b:'S -> 'b:'a:'S!, !DUP :: 'a:'S -> 'a:'a:'S!,
and !DROP :: 'a:'S -> 'S! have to be used to respectively change the
order of the values on the Michelson stack, to duplicate a value, and
to pop a value from the stack without actually using it.  Some
instructions, like comparison or arithmetic operations, exhibit
non-ambiguous ad-hoc polymorphism: depending on the input arguments'
type, a specific implementation of the instruction is selected, and
the return type is fixed. For example !SIZE! has the following types:

\begin{tabular}[t]{lcl}
\begin{lstlisting}
bytes:'S -> nat:'S
string:'S -> nat:'S
\end{lstlisting}
&\hspace{3em} &
\begin{lstlisting}
set 'elt:'S -> nat:'S
map 'key 'val:'S -> nat:'S
list 'elt:'S -> nat:'S
\end{lstlisting}
\end{tabular}

While computing the size of a string or an array of bytes is similarly implemented, under the hood,
the computation of map size has nothing to do with the computation of
string size.


Finally, the contract's code is required to take a stack
with a pair \emph{parameter}-\emph{storage}
and returns a stack with a pair \emph{operation list}-\emph{storage}:\\
!(parameter_ty*storage_ty):[] -> (operation list*storage_ty):[]!.

The operations listed at the end of the execution can change the
delegate of the contract, originate new contracts, or transfer tokens
to other addresses.  They will be executed right after the execution
of the contract.  The transfers can have parameters and trigger the
execution of other smart contracts: this is the only way to perform
\emph{inter-contract} calls.

\section{The Albert intermediate language}
\label{sec:albert}
\label{sec:albert_lang}

Michelson, as a stack-based language, is a difficult and unusual
target for compiler writers. In addition to the usual effort to
translate high-level constructions to lower-level types and
control-flow, they have to deal with stack manipulation to make values
available at the right stack position when calling an Michelson
opcode, and to cope with the consumption of values by the opcode
execution.

These additional difficulties also hinder the effort of teams
developing static analysers and  verification frameworks.

As a first simplification step, we have decided to build an
intermediate language that abstracts away the ordering of values in
the stack and provides a named binding to values.  This intermediate
language still keeps track of the resources as variables are typed by
a linear type system, which enforces each value to be consumed exactly
once. When a value is needed more than once, it must be explicitly
duplicated with a ?dup? operation. Generation of ?dup?s is left to a
future higher-level intermediate language.

In the process of defining the language, we thought that it would also
be helpful to define some abstractions over the datatypes so we
provide support for \emph{records} which compile to nestings of
Michelson's binary product type !pair! and \emph{variants} which
compile to nestings of Michelson's binary sum type !or!.

We also offer to define separate non-recursive function definitions
used to define programming libraries. These functions are
inlined at compile time.

\newcommand{\basedrule}[4][]{{\displaystyle\frac{\begin{array}{l}#2\end{array}}{#3}\quad\basedrulename{#4}}}
\newcommand{\baseusedrule}[1]{\[#1\]}
\newcommand{\basepremise}[1]{ #1 \\}
\newenvironment{basedefnblock}[3][]{ \framebox{\mbox{#2}} \quad #3 \\[0pt]}{}
\newenvironment{basefundefnblock}[3][]{ \framebox{\mbox{#2}} \quad #3 \\[0pt]\begin{displaymath}\begin{array}{l}}{\end{array}\end{displaymath}}
\newcommand{\basefunclause}[2]{ #1 \equiv #2 \\}
\newcommand{\basent}[1]{\mathit{#1}}
\newcommand{\basemv}[1]{\mathit{#1}}
\newcommand{\basekw}[1]{\mathbf{#1}}
\newcommand{\basesym}[1]{#1}
\newcommand{\basecom}[1]{\text{#1}}
\newcommand{\basedrulename}[1]{\textsc{#1}}
\newcommand{\basecomplu}[5]{\overline{#1}^{\,#2\in #3 #4 #5}}
\newcommand{\basecompu}[3]{\overline{#1}^{\,#2<#3}}
\newcommand{\basecomp}[2]{\overline{#1}^{\,#2}}
\newcommand{\basegrammartabular}[1]{\begin{supertabular}{llcllllll}#1\end{supertabular}}
\newcommand{\basemetavartabular}[1]{\begin{supertabular}{ll}#1\end{supertabular}}
\newcommand{\baserulehead}[3]{$#1$ & & $#2$ & & & \multicolumn{2}{l}{#3}}
\newcommand{\baseprodline}[6]{& & $#1$ & $#2$ & $#3 #4$ & $#5$ & $#6$}
\newcommand{\basefirstprodline}[6]{\baseprodline{#1}{#2}{#3}{#4}{#5}{#6}}
\newcommand{\baselongprodline}[2]{& & $#1$ & \multicolumn{4}{l}{$#2$}}
\newcommand{\basefirstlongprodline}[2]{\baselongprodline{#1}{#2}}
\newcommand{\basebindspecprodline}[6]{\baseprodline{#1}{#2}{#3}{#4}{#5}{#6}}
\newcommand{\baseprodnewline}{\\}
\newcommand{\baseinterrule}{\\[5.0mm]}
\newcommand{\baseafterlastrule}{\\}
\renewcommand{\basekw}{\texttt}

\newcommand{\basemetavars}{
\basemetavartabular{
 $ \basemv{n} ,\, \basemv{m} $ &  \\
 $ \basemv{id} $ &  \\
}}

\newcommand{\baselabel}{
\baserulehead{\basent{label}  ,\ \basent{l}  ,\ \basent{var}  ,\ \basent{x}}{::=}{\basecom{Label / variable}}\baseprodnewline
\basefirstprodline{|}{\basemv{id}}{}{}{}{}}

\newcommand{\baserty}{
\baserulehead{\basent{rty}}{::=}{\basecom{Record type}}\baseprodnewline
\basefirstprodline{|}{\lbracett  \basent{l_{{\mathrm{1}}}}  ~\texttt{:}~  \basent{ty_{{\mathrm{1}}}}  \texttt{;}~ \, .. \, \texttt{;}~  \basent{l_{\basemv{n}}}  ~\texttt{:}~  \basent{ty_{\basemv{n}}}  \rbracett}{}{}{}{}}

\newcommand{\basety}{
\baserulehead{\basent{ty}  ,\ \basent{a}  ,\ \basent{b}  ,\ \basent{c}}{::=}{\basecom{Type}}\baseprodnewline
\basefirstprodline{|}{\basent{rty}}{}{}{}{\basecom{Record type}}}

\newcommand{\baseinstruction}{
\baserulehead{\basent{instruction}  ,\ \basent{I}  ,\ \basent{ins}}{::=}{\basecom{Instruction}}\baseprodnewline
\basefirstprodline{|}{\basekw{noop}}{}{}{}{\basecom{No operation}}\baseprodnewline
\baseprodline{|}{\basent{instruction_{{\mathrm{1}}}}  \texttt{;}~  \basent{instruction_{{\mathrm{2}}}}}{}{}{}{\basecom{Sequencing}}\baseprodnewline
\baseprodline{|}{\basent{lhs}  \texttt{=}  \basent{rhs}}{}{}{}{\basecom{Assignment}}\baseprodnewline
\baseprodline{|}{\basekw{drop} \, \basent{var}}{}{}{}{\basecom{Resource dropping}}}

\newcommand{\baselhs}{
\baserulehead{\basent{lhs}}{::=}{\basecom{Left-hand side of assignement}}\baseprodnewline
\basefirstprodline{|}{\basent{var}}{}{}{}{}\baseprodnewline
\baseprodline{|}{\lbracett  \basent{l_{{\mathrm{1}}}}  \texttt{=}  \basent{var_{{\mathrm{1}}}}  \texttt{;}~ \, .. \, \texttt{;}~  \basent{l_{\basemv{n}}}  \texttt{=}  \basent{var_{\basemv{n}}}  \rbracett}{}{}{}{}}

\newcommand{\baserhs}{
\baserulehead{\basent{rhs}}{::=}{\basecom{Right-hand side of assignments}}\baseprodnewline
\basefirstprodline{|}{\basent{arg}}{}{}{}{}\baseprodnewline
\baseprodline{|}{\basent{f} \, \basent{arg}}{}{}{}{}\baseprodnewline
\baseprodline{|}{\basent{var}  \basesym{.}  \basent{l}}{}{}{}{}\baseprodnewline
\baseprodline{|}{\lbracett  \basent{var} \, \basekw{with} \, \basent{l_{{\mathrm{1}}}}  \texttt{=}  \basent{var_{{\mathrm{1}}}}  \texttt{;}~ \, ... \, \texttt{;}~  \basent{l_{\basemv{n}}}  \texttt{=}  \basent{var_{\basemv{n}}}  \rbracett}{}{}{}{}}

\newcommand{\basef}{
\baserulehead{\basent{f}}{::=}{\basecom{Function symbol}}\baseprodnewline
\basefirstprodline{|}{\basekw{dup}}{}{}{}{}}

\newcommand{\basearg}{
\baserulehead{\basent{arg}}{::=}{\basecom{Function argument}}\baseprodnewline
\basefirstprodline{|}{\basent{var}}{}{}{}{}\baseprodnewline
\baseprodline{|}{\basent{value}}{}{}{}{}\baseprodnewline
\baseprodline{|}{\lbracett  \basent{l_{{\mathrm{1}}}}  \texttt{=}  \basent{var_{{\mathrm{1}}}}  \texttt{;}~ \, ... \, \texttt{;}~  \basent{l_{\basemv{n}}}  \texttt{=}  \basent{var_{\basemv{n}}}  \rbracett}{}{}{}{}}

\newcommand{\baserval}{
\baserulehead{\basent{rval}}{::=}{\basecom{Record value}}\baseprodnewline
\basefirstprodline{|}{\lbracett  \basent{l_{{\mathrm{1}}}}  \texttt{=}  \basent{value_{{\mathrm{1}}}}  \texttt{;}~ \, .. \, \texttt{;}~  \basent{l_{\basemv{n}}}  \texttt{=}  \basent{value_{\basemv{n}}}  \rbracett}{}{}{}{\basecom{Record}}}

\newcommand{\basevalue}{
\baserulehead{\basent{value}  ,\ \basent{val}}{::=}{\basecom{Value}}\baseprodnewline
\basefirstprodline{|}{\basent{rval}}{}{}{}{\basecom{Record}}\baseprodnewline
\baseprodline{|}{\texttt{(}  \basent{value}  \texttt{)}}{}{}{}{}}

\newcommand{\baseg}{
\baserulehead{\Gamma}{::=}{\basecom{Typing context}}\baseprodnewline
\basefirstprodline{|}{\basesym{.}}{}{}{}{}}

\newcommand{\baseproducts}{
\baserulehead{\basent{products}}{::=}{\basecom{Products of types}}\baseprodnewline
\basefirstprodline{|}{\lbracett  \basent{ty_{{\mathrm{1}}}}  ~\texttt{*}~  \basent{ty'_{{\mathrm{1}}}}  \texttt{;}~ \, .. \, \texttt{;}~  \basent{ty_{\basemv{n}}}  ~\texttt{*}~  \basent{ty'_{\basemv{n}}}  \rbracett}{}{}{}{}}

\newcommand{\basegrammar}{\basegrammartabular{
\baselabel\baseinterrule
\baserty\baseinterrule
\basety\baseinterrule
\baseinstruction\baseinterrule
\baselhs\baseinterrule
\baserhs\baseinterrule
\basef\baseinterrule
\basearg\baseinterrule
\baserval\baseinterrule
\basevalue\baseinterrule
\baseg\baseinterrule
\baseproducts\baseafterlastrule
}}

\newcommand{\basedruletypingXXTeqXXrefl}[1]{\basedrule[#1]{%
}{
\Gamma  \vdash  \basent{ty}  \equiv  \basent{ty}}{%
{\basedrulename{typing\_Teq\_refl}}{}%
}}

\newcommand{\basedruletypingXXTeqXXsym}[1]{\basedrule[#1]{%
\basepremise{\Gamma  \vdash  \basent{ty_{{\mathrm{1}}}}  \equiv  \basent{ty_{{\mathrm{2}}}}}%
}{
\Gamma  \vdash  \basent{ty_{{\mathrm{2}}}}  \equiv  \basent{ty_{{\mathrm{1}}}}}{%
{\basedrulename{typing\_Teq\_sym}}{}%
}}

\newcommand{\basedruletypingXXTeqXXtrans}[1]{\basedrule[#1]{%
\basepremise{\Gamma  \vdash  \basent{ty_{{\mathrm{1}}}}  \equiv  \basent{ty_{{\mathrm{2}}}}}%
\basepremise{\Gamma  \vdash  \basent{ty_{{\mathrm{2}}}}  \equiv  \basent{ty_{{\mathrm{3}}}}}%
}{
\Gamma  \vdash  \basent{ty_{{\mathrm{1}}}}  \equiv  \basent{ty_{{\mathrm{3}}}}}{%
{\basedrulename{typing\_Teq\_trans}}{}%
}}

\newcommand{\basedruletypingXXTeqXXcongr}[1]{\basedrule[#1]{%
\basepremise{\Gamma  \vdash  \basent{ty_{{\mathrm{1}}}}  \equiv  \basent{ty'_{{\mathrm{1}}}} \quad .. \quad \Gamma  \vdash  \basent{ty_{\basemv{n}}}  \equiv  \basent{ty'_{\basemv{n}}}}%
}{
\Gamma  \vdash  \lbracett  \basent{l_{{\mathrm{1}}}}  ~\texttt{:}~  \basent{ty_{{\mathrm{1}}}}  \texttt{;}~ \, .. \, \texttt{;}~  \basent{l_{\basemv{n}}}  ~\texttt{:}~  \basent{ty_{\basemv{n}}}  \rbracett  \equiv  \lbracett  \basent{l_{{\mathrm{1}}}}  ~\texttt{:}~  \basent{ty'_{{\mathrm{1}}}}  \texttt{;}~ \, .. \, \texttt{;}~  \basent{l_{\basemv{n}}}  ~\texttt{:}~  \basent{ty'_{\basemv{n}}}  \rbracett}{%
{\basedrulename{typing\_Teq\_congr}}{}%
}}

\newcommand{\basedefntypingXXtypeXXeq}[1]{\begin{basedefnblock}[#1]{$\Gamma  \vdash  \basent{ty_{{\mathrm{1}}}}  \equiv  \basent{ty_{{\mathrm{2}}}}$}{\basecom{Type equality}}
\baseusedrule{\basedruletypingXXTeqXXrefl{}}
\baseusedrule{\basedruletypingXXTeqXXsym{}}
\baseusedrule{\basedruletypingXXTeqXXtrans{}}
\baseusedrule{\basedruletypingXXTeqXXcongr{}}
\end{basedefnblock}}

\newcommand{\basedruletypingXXtwfXXrecord}[1]{\basedrule[#1]{%
\basepremise{ l_1 < .. < l_n }%
\basepremise{\Gamma  \vdash  \basent{ty_{{\mathrm{1}}}} \quad .. \quad \Gamma  \vdash  \basent{ty_{\basemv{n}}}}%
\basepremise{\Gamma  \vdash}%
}{
\Gamma  \vdash  \lbracett  \basent{l_{{\mathrm{1}}}}  ~\texttt{:}~  \basent{ty_{{\mathrm{1}}}}  \texttt{;}~ \, .. \, \texttt{;}~  \basent{l_{\basemv{n}}}  ~\texttt{:}~  \basent{ty_{\basemv{n}}}  \rbracett}{%
{\basedrulename{typing\_twf\_record}}{}%
}}

\newcommand{\basedefntypingXXtyXXwellXXformed}[1]{\begin{basedefnblock}[#1]{$\Gamma  \vdash  \basent{ty}$}{\basecom{Type well-formedness}}
\baseusedrule{\basedruletypingXXtwfXXrecord{}}
\end{basedefnblock}}

\newcommand{\basedruletypingXXGwfXXempty}[1]{\basedrule[#1]{%
}{
\basesym{.}  \vdash}{%
{\basedrulename{typing\_Gwf\_empty}}{}%
}}

\newcommand{\basedefntypingXXcontextXXwellXXformed}[1]{\begin{basedefnblock}[#1]{$\Gamma  \vdash$}{\basecom{Context well-fornedness}}
\baseusedrule{\basedruletypingXXGwfXXempty{}}
\end{basedefnblock}}

\newcommand{\basedruletypingXXTfreshXXtvarrecord}[1]{\basedrule[#1]{%
\basepremise{ \basekw{tvar}  \notin \mathsf{FV}( \basent{ty_{{\mathrm{1}}}} )  \quad .. \quad  \basekw{tvar}  \notin \mathsf{FV}( \basent{ty_{\basemv{n}}} ) }%
}{
 \basekw{tvar}  \notin \mathsf{FV}( \lbracett  \basent{l_{{\mathrm{1}}}}  ~\texttt{:}~  \basent{ty_{{\mathrm{1}}}}  \texttt{;}~ \, .. \, \texttt{;}~  \basent{l_{\basemv{n}}}  ~\texttt{:}~  \basent{ty_{\basemv{n}}}  \rbracett ) }{%
{\basedrulename{typing\_Tfresh\_tvarrecord}}{}%
}}

\newcommand{\basedefntypingXXfreshXXtvar}[1]{\begin{basedefnblock}[#1]{$ \basekw{tvar}  \notin \mathsf{FV}( \basent{ty} ) $}{\basecom{Type variable freshness}}
\baseusedrule{\basedruletypingXXTfreshXXtvarrecord{}}
\end{basedefnblock}}

\newcommand{\basedruletypingXXTXXframe}[1]{\basedrule[#1]{%
\basepremise{\Gamma  \vdash  \basent{rty_{{\mathrm{1}}}}}%
\basepremise{\Gamma  \vdash  \basent{rty_{{\mathrm{2}}}}}%
\basepremise{\basent{rty}  \odot  \basent{rty''}  \texttt{=}  \basent{rty_{{\mathrm{1}}}}}%
\basepremise{\basent{rty'}  \odot  \basent{rty''}  \texttt{=}  \basent{rty_{{\mathrm{2}}}}}%
\basepremise{\Gamma  \vdash  \basent{instruction}  ~\texttt{:}~  \basent{rty}  \Rightarrow  \basent{rty'}}%
}{
\Gamma  \vdash  \basent{instruction}  ~\texttt{:}~  \basent{rty_{{\mathrm{1}}}}  \Rightarrow  \basent{rty_{{\mathrm{2}}}}}{%
{\basedrulename{typing\_T\_frame}}{}%
}}

\newcommand{\basedruletypingXXTXXnoop}[1]{\basedrule[#1]{%
}{
\Gamma  \vdash  \basekw{noop}  ~\texttt{:}~  \lbracett  \,  \rbracett  \Rightarrow  \lbracett  \,  \rbracett}{%
{\basedrulename{typing\_T\_noop}}{}%
}}

\newcommand{\basedruletypingXXTXXseq}[1]{\basedrule[#1]{%
\basepremise{\Gamma  \vdash  \basent{instruction}  ~\texttt{:}~  \basent{ty_{{\mathrm{1}}}}  \Rightarrow  \basent{ty_{{\mathrm{2}}}}}%
\basepremise{\Gamma  \vdash  \basent{instruction'}  ~\texttt{:}~  \basent{ty_{{\mathrm{2}}}}  \Rightarrow  \basent{ty_{{\mathrm{3}}}}}%
}{
\Gamma  \vdash  \basent{instruction}  \texttt{;}~  \basent{instruction'}  ~\texttt{:}~  \basent{ty_{{\mathrm{1}}}}  \Rightarrow  \basent{ty_{{\mathrm{3}}}}}{%
{\basedrulename{typing\_T\_seq}}{}%
}}

\newcommand{\basedruletypingXXTXXassign}[1]{\basedrule[#1]{%
\basepremise{\Gamma  \vdash  \basent{rhs}  ~\texttt{:}~  \basent{a}  \Rightarrow  \basent{b}}%
\basepremise{\Gamma  \vdash  \basent{lhs}  ~\texttt{:}~  \basent{b}  \Rightarrow  \basent{c}}%
}{
\Gamma  \vdash  \basent{lhs}  \texttt{=}  \basent{rhs}  ~\texttt{:}~  \basent{a}  \Rightarrow  \basent{c}}{%
{\basedrulename{typing\_T\_assign}}{}%
}}

\newcommand{\basedruletypingXXTXXdrop}[1]{\basedrule[#1]{%
}{
\Gamma  \vdash  \basekw{drop} \, \basent{var}  ~\texttt{:}~  \lbracett  \basent{var}  ~\texttt{:}~  \basent{ty}  \rbracett  \Rightarrow  \lbracett  \,  \rbracett}{%
{\basedrulename{typing\_T\_drop}}{}%
}}

\newcommand{\basedefntypingXXinstr}[1]{\begin{basedefnblock}[#1]{$\Gamma  \vdash  \basent{instruction}  ~\texttt{:}~  \basent{ty}  \Rightarrow  \basent{ty'}$}{\basecom{Instruction typing}}
\baseusedrule{\basedruletypingXXTXXframe{}}
\baseusedrule{\basedruletypingXXTXXnoop{}}
\baseusedrule{\basedruletypingXXTXXseq{}}
\baseusedrule{\basedruletypingXXTXXassign{}}
\baseusedrule{\basedruletypingXXTXXdrop{}}
\end{basedefnblock}}

\newcommand{\basedruletypingXXTlhsXXvar}[1]{\basedrule[#1]{%
}{
\Gamma  \vdash  \basent{var}  ~\texttt{:}~  \basent{ty}  \Rightarrow  \lbracett  \basent{var}  ~\texttt{:}~  \basent{ty}  \rbracett}{%
{\basedrulename{typing\_Tlhs\_var}}{}%
}}

\newcommand{\basedruletypingXXTlhsXXrecord}[1]{\basedrule[#1]{%
}{
\Gamma  \vdash  \lbracett  \basent{l_{{\mathrm{1}}}}  \texttt{=}  \basent{x_{{\mathrm{1}}}}  \texttt{;}~ \, .. \, \texttt{;}~  \basent{l_{\basemv{n}}}  \texttt{=}  \basent{x_{\basemv{n}}}  \rbracett  ~\texttt{:}~  \lbracett  \basent{l_{{\mathrm{1}}}}  ~\texttt{:}~  \basent{ty_{{\mathrm{1}}}}  \texttt{;}~ \, .. \, \texttt{;}~  \basent{l_{\basemv{n}}}  ~\texttt{:}~  \basent{ty_{\basemv{n}}}  \rbracett  \Rightarrow  \lbracett  \basent{x_{{\mathrm{1}}}}  ~\texttt{:}~  \basent{ty_{{\mathrm{1}}}}  \texttt{;}~ \, .. \, \texttt{;}~  \basent{x_{\basemv{n}}}  ~\texttt{:}~  \basent{ty_{\basemv{n}}}  \rbracett}{%
{\basedrulename{typing\_Tlhs\_record}}{}%
}}

\newcommand{\basedefntypingXXlhs}[1]{\begin{basedefnblock}[#1]{$\Gamma  \vdash  \basent{lhs}  ~\texttt{:}~  \basent{ty}  \Rightarrow  \basent{ty'}$}{\basecom{Left-hand sides typing}}
\baseusedrule{\basedruletypingXXTlhsXXvar{}}
\baseusedrule{\basedruletypingXXTlhsXXrecord{}}
\end{basedefnblock}}

\newcommand{\basedruletypingXXTargXXvar}[1]{\basedrule[#1]{%
}{
\Gamma  \vdash_a  \basent{var}  ~\texttt{:}~  \lbracett  \basent{var}  ~\texttt{:}~  \basent{ty}  \rbracett  \Rightarrow  \basent{ty}}{%
{\basedrulename{typing\_Targ\_var}}{}%
}}

\newcommand{\basedruletypingXXTargXXvalue}[1]{\basedrule[#1]{%
\basepremise{\Gamma  \vdash  \basent{value}  ~\texttt{:}~  \basent{ty}}%
}{
\Gamma  \vdash_a  \basent{value}  ~\texttt{:}~  \lbracett  \,  \rbracett  \Rightarrow  \basent{ty}}{%
{\basedrulename{typing\_Targ\_value}}{}%
}}

\newcommand{\basedruletypingXXTargXXrecord}[1]{\basedrule[#1]{%
}{
\Gamma  \vdash_a  \lbracett  \basent{l_{{\mathrm{1}}}}  \texttt{=}  \basent{x_{{\mathrm{1}}}}  \texttt{;}~ \, .. \, \texttt{;}~  \basent{l_{\basemv{n}}}  \texttt{=}  \basent{x_{\basemv{n}}}  \rbracett  ~\texttt{:}~  \lbracett  \basent{x_{{\mathrm{1}}}}  ~\texttt{:}~  \basent{ty_{{\mathrm{1}}}}  \texttt{;}~ \, .. \, \texttt{;}~  \basent{x_{\basemv{n}}}  ~\texttt{:}~  \basent{ty_{\basemv{n}}}  \rbracett  \Rightarrow  \lbracett  \basent{l_{{\mathrm{1}}}}  ~\texttt{:}~  \basent{ty_{{\mathrm{1}}}}  \texttt{;}~ \, .. \, \texttt{;}~  \basent{l_{\basemv{n}}}  ~\texttt{:}~  \basent{ty_{\basemv{n}}}  \rbracett}{%
{\basedrulename{typing\_Targ\_record}}{}%
}}

\newcommand{\basedefntypingXXarg}[1]{\begin{basedefnblock}[#1]{$\Gamma  \vdash_a  \basent{arg}  ~\texttt{:}~  \basent{ty}  \Rightarrow  \basent{ty'}$}{\basecom{Argument typing}}
\baseusedrule{\basedruletypingXXTargXXvar{}}
\baseusedrule{\basedruletypingXXTargXXvalue{}}
\baseusedrule{\basedruletypingXXTargXXrecord{}}
\end{basedefnblock}}

\newcommand{\basedruletypingXXTrhsXXarg}[1]{\basedrule[#1]{%
\basepremise{\Gamma  \vdash_a  \basent{arg}  ~\texttt{:}~  \basent{ty}  \Rightarrow  \basent{ty'}}%
}{
\Gamma  \vdash  \basent{arg}  ~\texttt{:}~  \basent{ty}  \Rightarrow  \basent{ty'}}{%
{\basedrulename{typing\_Trhs\_arg}}{}%
}}

\newcommand{\basedruletypingXXTrhsXXf}[1]{\basedrule[#1]{%
\basepremise{\Gamma  \vdash_a  \basent{arg}  ~\texttt{:}~  \basent{ty}  \Rightarrow  \basent{ty'}}%
\basepremise{\Gamma  \vdash  \basent{f}  ~\texttt{:}~  \basent{ty'}  \Rightarrow  \basent{ty''}}%
}{
\Gamma  \vdash  \basent{f} \, \basent{arg}  ~\texttt{:}~  \basent{ty}  \Rightarrow  \basent{ty''}}{%
{\basedrulename{typing\_Trhs\_f}}{}%
}}

\newcommand{\basedruletypingXXTrhsXXprojection}[1]{\basedrule[#1]{%
\basepremise{\lbracett  \basent{l}  ~\texttt{:}~  \basent{ty}  \rbracett  \odot  \basent{rty}  \texttt{=}  \basent{rty'}}%
\basepremise{\Gamma  \vdash  \basent{rty'}}%
}{
\Gamma  \vdash  \basent{var}  \basesym{.}  \basent{l}  ~\texttt{:}~  \basent{rty'}  \Rightarrow  \basent{ty}}{%
{\basedrulename{typing\_Trhs\_projection}}{}%
}}

\newcommand{\basedruletypingXXTrhsXXupdate}[1]{\basedrule[#1]{%
\basepremise{\Gamma  \vdash  \basent{rty'}}%
\basepremise{\lbracett  \basent{l_{{\mathrm{1}}}}  ~\texttt{:}~  \basent{ty_{{\mathrm{1}}}}  \texttt{;}~ \, .. \, \texttt{;}~  \basent{l_{\basemv{n}}}  ~\texttt{:}~  \basent{ty_{\basemv{n}}}  \rbracett  \odot  \basent{rty}  \texttt{=}  \basent{rty'}}%
}{
\Gamma  \vdash  \lbracett  \basent{var} \, \basekw{with} \, \basent{l_{{\mathrm{1}}}}  \texttt{=}  \basent{var_{{\mathrm{1}}}}  \texttt{;}~ \, .. \, \texttt{;}~  \basent{l_{\basemv{n}}}  \texttt{=}  \basent{var_{\basemv{n}}}  \rbracett  ~\texttt{:}~  \lbracett  \basent{var}  ~\texttt{:}~  \basent{rty'}  \texttt{;}~  \basent{var_{{\mathrm{1}}}}  ~\texttt{:}~  \basent{ty_{{\mathrm{1}}}}  \texttt{;}~ \, .. \, \texttt{;}~  \basent{var_{\basemv{n}}}  ~\texttt{:}~  \basent{ty_{\basemv{n}}}  \rbracett  \Rightarrow  \basent{rty'}}{%
{\basedrulename{typing\_Trhs\_update}}{}%
}}

\newcommand{\basedefntypingXXrhs}[1]{\begin{basedefnblock}[#1]{$\Gamma  \vdash  \basent{rhs}  ~\texttt{:}~  \basent{ty}  \Rightarrow  \basent{ty'}$}{\basecom{Right-hand side typing}}
\baseusedrule{\basedruletypingXXTrhsXXarg{}}
\baseusedrule{\basedruletypingXXTrhsXXf{}}
\baseusedrule{\basedruletypingXXTrhsXXprojection{}}
\baseusedrule{\basedruletypingXXTrhsXXupdate{}}
\end{basedefnblock}}

\newcommand{\basedruletypingXXTfXXdup}[1]{\basedrule[#1]{%
}{
\Gamma  \vdash  \basekw{dup}  ~\texttt{:}~  \basent{ty}  \Rightarrow  \lbracett  \basekw{car}  ~\texttt{:}~  \basent{ty}  \texttt{;}~  \basekw{cdr}  ~\texttt{:}~  \basent{ty}  \rbracett}{%
{\basedrulename{typing\_Tf\_dup}}{}%
}}

\newcommand{\basedefntypingXXf}[1]{\begin{basedefnblock}[#1]{$\Gamma  \vdash  \basent{f}  ~\texttt{:}~  \basent{ty}  \Rightarrow  \basent{ty'}$}{\basecom{Function symbol typing}}
\baseusedrule{\basedruletypingXXTfXXdup{}}
\end{basedefnblock}}

\newcommand{\basedruletypingXXTvalXXrecord}[1]{\basedrule[#1]{%
\basepremise{\Gamma  \vdash  \basent{val_{{\mathrm{1}}}}  ~\texttt{:}~  \basent{ty_{{\mathrm{1}}}} \quad .. \quad \Gamma  \vdash  \basent{val_{\basemv{n}}}  ~\texttt{:}~  \basent{ty_{\basemv{n}}}}%
}{
\Gamma  \vdash  \lbracett  \basent{l_{{\mathrm{1}}}}  \texttt{=}  \basent{val_{{\mathrm{1}}}}  \texttt{;}~ \, .. \, \texttt{;}~  \basent{l_{\basemv{n}}}  \texttt{=}  \basent{val_{\basemv{n}}}  \rbracett  ~\texttt{:}~  \lbracett  \basent{l_{{\mathrm{1}}}}  ~\texttt{:}~  \basent{ty_{{\mathrm{1}}}}  \texttt{;}~ \, .. \, \texttt{;}~  \basent{l_{\basemv{n}}}  ~\texttt{:}~  \basent{ty_{\basemv{n}}}  \rbracett}{%
{\basedrulename{typing\_Tval\_record}}{}%
}}

\newcommand{\basedefntypingXXval}[1]{\begin{basedefnblock}[#1]{$\Gamma  \vdash  \basent{value}  ~\texttt{:}~  \basent{ty}$}{\basecom{Value typing}}
\baseusedrule{\basedruletypingXXTvalXXrecord{}}
\end{basedefnblock}}

\newcommand{\basedefnsJtype}{
\basedefntypingXXtypeXXeq{}\basedefntypingXXtyXXwellXXformed{}\basedefntypingXXcontextXXwellXXformed{}\basedefntypingXXfreshXXtvar{}\basedefntypingXXinstr{}\basedefntypingXXlhs{}\basedefntypingXXarg{}\basedefntypingXXrhs{}\basedefntypingXXf{}\basedefntypingXXval{}}

\newcommand{\basedruleevalXXlhsXXvar}[1]{\basedrule[#1]{%
}{
\basent{var}  \basesym{/}  \basent{val}  ~\Longrightarrow~  \lbracett  \basent{var}  \texttt{=}  \basent{val}  \rbracett}{%
{\basedrulename{eval\_lhs\_var}}{}%
}}

\newcommand{\basedruleevalXXlhsXXrecord}[1]{\basedrule[#1]{%
}{
\lbracett  \basent{l_{{\mathrm{1}}}}  \texttt{=}  \basent{x_{{\mathrm{1}}}}  \texttt{;}~ \, .. \, \texttt{;}~  \basent{l_{\basemv{n}}}  \texttt{=}  \basent{x_{\basemv{n}}}  \rbracett  \basesym{/}  \lbracett  \basent{l_{{\mathrm{1}}}}  \texttt{=}  \basent{val_{{\mathrm{1}}}}  \texttt{;}~ \, .. \, \texttt{;}~  \basent{l_{\basemv{n}}}  \texttt{=}  \basent{val_{\basemv{n}}}  \rbracett  ~\Longrightarrow~  \lbracett  \basent{x_{{\mathrm{1}}}}  \texttt{=}  \basent{val_{{\mathrm{1}}}}  \texttt{;}~ \, .. \, \texttt{;}~  \basent{x_{\basemv{n}}}  \texttt{=}  \basent{val_{\basemv{n}}}  \rbracett}{%
{\basedrulename{eval\_lhs\_record}}{}%
}}

\newcommand{\basedefnevalXXlhs}[1]{\begin{basedefnblock}[#1]{$\basent{lhs}  \basesym{/}  \basent{val}  ~\Longrightarrow~  \basent{val'}$}{\basecom{Left-hand side evaluation}}
\baseusedrule{\basedruleevalXXlhsXXvar{}}
\baseusedrule{\basedruleevalXXlhsXXrecord{}}
\end{basedefnblock}}

\newcommand{\basedruleevalXXargXXvar}[1]{\basedrule[#1]{%
}{
\basent{var}  /_a  \lbracett  \basent{var}  \texttt{=}  \basent{val}  \rbracett  ~\Longrightarrow~  \basent{val}}{%
{\basedrulename{eval\_arg\_var}}{}%
}}

\newcommand{\basedruleevalXXargXXval}[1]{\basedrule[#1]{%
}{
\basent{val}  /_a  \lbracett  \,  \rbracett  ~\Longrightarrow~  \basent{val}}{%
{\basedrulename{eval\_arg\_val}}{}%
}}

\newcommand{\basedruleevalXXargXXrecord}[1]{\basedrule[#1]{%
}{
\lbracett  \basent{l_{{\mathrm{1}}}}  \texttt{=}  \basent{x_{{\mathrm{1}}}}  \texttt{;}~ \, .. \, \texttt{;}~  \basent{l_{\basemv{n}}}  \texttt{=}  \basent{x_{\basemv{n}}}  \rbracett  /_a  \lbracett  \basent{x_{{\mathrm{1}}}}  \texttt{=}  \basent{val_{{\mathrm{1}}}}  \texttt{;}~ \, .. \, \texttt{;}~  \basent{x_{\basemv{n}}}  \texttt{=}  \basent{val_{\basemv{n}}}  \rbracett  ~\Longrightarrow~  \lbracett  \basent{l_{{\mathrm{1}}}}  \texttt{=}  \basent{val_{{\mathrm{1}}}}  \texttt{;}~ \, .. \, \texttt{;}~  \basent{l_{\basemv{n}}}  \texttt{=}  \basent{val_{\basemv{n}}}  \rbracett}{%
{\basedrulename{eval\_arg\_record}}{}%
}}

\newcommand{\basedefnevalXXarg}[1]{\begin{basedefnblock}[#1]{$\basent{arg}  /_a  \basent{val}  ~\Longrightarrow~  \basent{val'}$}{\basecom{Argument evaluation}}
\baseusedrule{\basedruleevalXXargXXvar{}}
\baseusedrule{\basedruleevalXXargXXval{}}
\baseusedrule{\basedruleevalXXargXXrecord{}}
\end{basedefnblock}}

\newcommand{\basedruleevalXXfXXdup}[1]{\basedrule[#1]{%
}{
\basekw{dup}  \basesym{/}  \basent{val}  ~\Longrightarrow~  \lbracett  \basekw{car}  \texttt{=}  \basent{val}  \texttt{;}~  \basekw{cdr}  \texttt{=}  \basent{val}  \rbracett}{%
{\basedrulename{eval\_f\_dup}}{}%
}}

\newcommand{\basedefnevalXXf}[1]{\begin{basedefnblock}[#1]{$\basent{f}  \basesym{/}  \basent{val}  ~\Longrightarrow~  \basent{val'}$}{\basecom{Function symbol evaluation}}
\baseusedrule{\basedruleevalXXfXXdup{}}
\end{basedefnblock}}

\newcommand{\basedruleevalXXrhsXXarg}[1]{\basedrule[#1]{%
\basepremise{\basent{arg}  /_a  \basent{val}  ~\Longrightarrow~  \basent{val'}}%
}{
\basent{arg}  \basesym{/}  \basent{val}  ~\Longrightarrow~  \basent{val'}}{%
{\basedrulename{eval\_rhs\_arg}}{}%
}}

\newcommand{\basedruleevalXXrhsXXf}[1]{\basedrule[#1]{%
\basepremise{\basent{arg}  /_a  \basent{val}  ~\Longrightarrow~  \basent{val'}}%
\basepremise{\basent{f}  \basesym{/}  \basent{val'}  ~\Longrightarrow~  \basent{val''}}%
}{
\basent{f} \, \basent{arg}  \basesym{/}  \basent{val}  ~\Longrightarrow~  \basent{val''}}{%
{\basedrulename{eval\_rhs\_f}}{}%
}}

\newcommand{\basedruleevalXXrhsXXprojection}[1]{\basedrule[#1]{%
\basepremise{\lbracett  \basent{l}  \texttt{=}  \basent{val}  \rbracett  \odot  \basent{rval}  \texttt{=}  \basent{rval'}}%
}{
\basent{var}  \basesym{.}  \basent{l}  \basesym{/}  \basent{rval'}  ~\Longrightarrow~  \basent{val}}{%
{\basedrulename{eval\_rhs\_projection}}{}%
}}

\newcommand{\basedruleevalXXrhsXXupdate}[1]{\basedrule[#1]{%
\basepremise{\lbracett  \basent{l_{{\mathrm{1}}}}  \texttt{=}  \basent{val'_{{\mathrm{1}}}}  \texttt{;}~ \, .. \, \texttt{;}~  \basent{l_{\basemv{n}}}  \texttt{=}  \basent{val'_{\basemv{n}}}  \rbracett  \odot  \basent{rval}  \texttt{=}  \basent{rval'}}%
\basepremise{\lbracett  \basent{l_{{\mathrm{1}}}}  \texttt{=}  \basent{val_{{\mathrm{1}}}}  \texttt{;}~ \, .. \, \texttt{;}~  \basent{l_{\basemv{n}}}  \texttt{=}  \basent{val_{\basemv{n}}}  \rbracett  \odot  \basent{rval}  \texttt{=}  \basent{rval''}}%
}{
\lbracett  \basent{var} \, \basekw{with} \, \basent{l_{{\mathrm{1}}}}  \texttt{=}  \basent{var_{{\mathrm{1}}}}  \texttt{;}~ \, .. \, \texttt{;}~  \basent{l_{\basemv{n}}}  \texttt{=}  \basent{var_{\basemv{n}}}  \rbracett  \basesym{/}  \lbracett  \basent{var}  \texttt{=}  \basent{rval'}  \texttt{;}~  \basent{var_{{\mathrm{1}}}}  \texttt{=}  \basent{val_{{\mathrm{1}}}}  \texttt{;}~ \, .. \, \texttt{;}~  \basent{var_{\basemv{n}}}  \texttt{=}  \basent{val_{\basemv{n}}}  \rbracett  ~\Longrightarrow~  \basent{rval''}}{%
{\basedrulename{eval\_rhs\_update}}{}%
}}

\newcommand{\basedefnevalXXrhs}[1]{\begin{basedefnblock}[#1]{$\basent{rhs}  \basesym{/}  \basent{val}  ~\Longrightarrow~  \basent{val'}$}{\basecom{Right-hand side evaluation}}
\baseusedrule{\basedruleevalXXrhsXXarg{}}
\baseusedrule{\basedruleevalXXrhsXXf{}}
\baseusedrule{\basedruleevalXXrhsXXprojection{}}
\baseusedrule{\basedruleevalXXrhsXXupdate{}}
\end{basedefnblock}}

\newcommand{\basedruleevalXXinstrXXframe}[1]{\basedrule[#1]{%
\basepremise{\basent{instruction}  \basesym{/}  \basent{rval}  ~\Longrightarrow~  \basent{rval'}}%
\basepremise{\basent{rval}  \odot  \basent{rval''}  \texttt{=}  \basent{rval_{{\mathrm{1}}}}}%
\basepremise{\basent{rval'}  \odot  \basent{rval''}  \texttt{=}  \basent{rval_{{\mathrm{2}}}}}%
}{
\basent{instruction}  \basesym{/}  \basent{rval_{{\mathrm{1}}}}  ~\Longrightarrow~  \basent{rval_{{\mathrm{2}}}}}{%
{\basedrulename{eval\_instr\_frame}}{}%
}}

\newcommand{\basedruleevalXXinstrXXnoop}[1]{\basedrule[#1]{%
}{
\basekw{noop}  \basesym{/}  \lbracett  \,  \rbracett  ~\Longrightarrow~  \lbracett  \,  \rbracett}{%
{\basedrulename{eval\_instr\_noop}}{}%
}}

\newcommand{\basedruleevalXXinstrXXseq}[1]{\basedrule[#1]{%
\basepremise{\basent{I_{{\mathrm{1}}}}  \basesym{/}  \basent{val}  ~\Longrightarrow~  \basent{val'}}%
\basepremise{\basent{I_{{\mathrm{2}}}}  \basesym{/}  \basent{val'}  ~\Longrightarrow~  \basent{val''}}%
}{
\basent{I_{{\mathrm{1}}}}  \texttt{;}~  \basent{I_{{\mathrm{2}}}}  \basesym{/}  \basent{val}  ~\Longrightarrow~  \basent{val''}}{%
{\basedrulename{eval\_instr\_seq}}{}%
}}

\newcommand{\basedruleevalXXinstrXXassign}[1]{\basedrule[#1]{%
\basepremise{\basent{rhs}  \basesym{/}  \basent{val}  ~\Longrightarrow~  \basent{val'}}%
\basepremise{\basent{lhs}  \basesym{/}  \basent{val'}  ~\Longrightarrow~  \basent{val''}}%
}{
\basent{lhs}  \texttt{=}  \basent{rhs}  \basesym{/}  \basent{val}  ~\Longrightarrow~  \basent{val''}}{%
{\basedrulename{eval\_instr\_assign}}{}%
}}

\newcommand{\basedruleevalXXinstrXXdrop}[1]{\basedrule[#1]{%
}{
\basekw{drop} \, \basent{var}  \basesym{/}  \lbracett  \basent{var}  \texttt{=}  \basent{val}  \rbracett  ~\Longrightarrow~  \lbracett  \,  \rbracett}{%
{\basedrulename{eval\_instr\_drop}}{}%
}}

\newcommand{\basedefnevalXXinstr}[1]{\begin{basedefnblock}[#1]{$\basent{instruction}  \basesym{/}  \basent{val}  ~\Longrightarrow~  \basent{val'}$}{\basecom{Instruction evaluation}}
\baseusedrule{\basedruleevalXXinstrXXframe{}}
\baseusedrule{\basedruleevalXXinstrXXnoop{}}
\baseusedrule{\basedruleevalXXinstrXXseq{}}
\baseusedrule{\basedruleevalXXinstrXXassign{}}
\baseusedrule{\basedruleevalXXinstrXXdrop{}}
\end{basedefnblock}}

\newcommand{\basedefnsJeval}{
\basedefnevalXXlhs{}\basedefnevalXXarg{}\basedefnevalXXf{}\basedefnevalXXrhs{}\basedefnevalXXinstr{}}

\newcommand{\basedefnss}{
\basedefnsJtype
\basedefnsJeval
}

\newcommand{\baseall}{\basemetavars\\[0pt]
\basegrammar\\[5.0mm]
\basedefnss}

\renewcommand{\basedrulename}[1]{}
\renewcommand{\basedrule}[4][]{{\displaystyle\frac{\begin{array}{l}#2\end{array}}{#3}\quad{#4}}}
\renewcommand{\basedruletypingXXTXXframe}[1]{\basedrule[#1]{%
\basepremise{\Gamma  \vdash  \basent{rty_{{\mathrm{1}}}}}%
\basepremise{\Gamma  \vdash  \basent{rty_{{\mathrm{2}}}}}%
\basepremise{\basent{rty}  \odot  \basent{rty''}  \texttt{=}  \basent{rty_{{\mathrm{1}}}}}%
\basepremise{\basent{rty'}  \odot  \basent{rty''}  \texttt{=}  \basent{rty_{{\mathrm{2}}}}}%
\basepremise{\Gamma  \vdash  \basent{instruction}  ~\texttt{:}~  \basent{rty}  \Rightarrow  \basent{rty'}}%
}{
\Gamma  \vdash  \basent{instruction}  ~\texttt{:}~  \basent{rty_{{\mathrm{1}}}}  \Rightarrow  \basent{rty_{{\mathrm{2}}}}}{%
{\textsc{Frame}}{}%
}
}

\subsection{Base language}

The Albert language is defined as a collection of small language
fragments that can be studied independently. Each fragment is defined
in a separate Ott file. The first fragment to
consider is called the \textit{base} fragment. As its name suggests,
this fragment is the basis on top of which the other fragments are
defined.

The base fragment contains the two main features of Albert: the stack
is abstracted by named variables and Michelson binary pairs are
generalized as records. We use the metavariable \(\basent{l}\) to denote
record labels and the metavariable \(\basent{x}\) to denote variables but these
two notions are unified in Albert.

\subsubsection{Records and linear typing}

As we have seen it Section~\ref{sec:instruction-type}, Michelson uses
an ordered type system that tracks both the order of the values on the
stack and the number of uses of the values. Most high-level languages
however bind values to named variables and implicitly handle the
ordering and number of uses of variables. The required stack
manipulation instructions are introduced at compile time. Albert is an
intermediate language between these two extremes. In Albert, the order
of values is abstracted but not the number of uses which is still
explicitly handled.

This choice is reflected in Albert's type system by the use of linear
typing. Each expression of the Albert language is typed by a pair of
record types whose labels are the variables touched by the instruction
or expression; the first record type describes the consumed values and
the second record type describes the produced values.

Thanks to the unification of variable names and record labels, records
in Albert generalize both the Michelson stack types and the Michelson
pair type. In the base fragment of Albert, all types are
possibly-empty record types.

The grammar of types of the base fragment given in Figure
\ref{fig:record-grammar}.
\renewcommand{\baselabel}{
\baserulehead{\basent{label}  ,\ \basent{l}}{::=}{\basecom{Label / variable}}\baseprodnewline
\basefirstprodline{|}{\basemv{id}}{}{}{}{}}

\renewcommand{\basety}{
\baserulehead{\basent{ty}}{::=}{\basecom{Type}}\baseprodnewline
\basefirstprodline{|}{\basent{rty}}{}{}{}{\basecom{Record type}}}

\begin{figure}
\centering  \begin{tabular}{l@{\hspace{5em}}r}
      \basegrammartabular{\baselabel\baseinterrule}&
      \basegrammartabular{\basety\baseinterrule} 
  \end{tabular}
    \basegrammartabular{\baserty\baseafterlastrule}
\caption{\label{fig:record-grammar}Syntax of the record types}
\end{figure}

In the record type \(\lbracett  \basent{l_{{\mathrm{1}}}}  ~\texttt{:}~  \basent{ty_{{\mathrm{1}}}}  \texttt{;}~ \, .. \, \texttt{;}~  \basent{l_{\basemv{n}}}  ~\texttt{:}~  \basent{ty_{\basemv{n}}}  \rbracett\), we
assume the labels to be distinct and lexicographically ordered.

This constraint is formalized by the well-formedness judgement
\(\Gamma  \vdash  \basent{ty}\) defined in Figure \ref{fig:record-wf}. The
typing context \(\Gamma\) is always empty here but other cases for
typing contexts will be added in other language fragments.


\begin{figure}
\basedefntypingXXtyXXwellXXformed{}
\caption{Type well-formedness judgment}
\label{fig:record-wf}
\end{figure}

The grammar for the base fragment is defined in Figure
\ref{fig:syntax-base}.  Albert's grammar is more stratified than
Michelson's grammar because we adopt from imperative languages the
usual distinction between expressions and instructions. An instruction
is either the \(\basekw{noop}\) instruction that does nothing, a sequence
of instructions separated by semicolons, or an assignment
\(\basent{lhs}  \texttt{=}  \basent{rhs}\) where the left-hand side \(\basent{lhs}\) is either a
variable or a record of variables and the right-hand side is an
expression.

Contrary to usual imperative expressions, arbitrary nesting of
expressions is not allowed and intermediate values should be named.
This restriction, inspired by the static single assignment form
commonly used in intermediate compilation languages, is designed to ease the
production of Michelson code and to allow for more optimisations at
the level of the Albert language in the future. In practice, this
restriction means that an expression is either a variable \(\basent{x}\), a
value \(\basent{val}\), the application of a user-defined function to a variable
\(\basent{f} \, \basent{x}\), a record projection \(\basent{x}  \basesym{.}  \basent{l}\), or a record update \(\lbracett  \basent{var} \, \basekw{with} \, \basent{l_{{\mathrm{1}}}}  \texttt{=}  \basent{var_{{\mathrm{1}}}}  \texttt{;}~ \, ... \, \texttt{;}~  \basent{l_{\basemv{n}}}  \texttt{=}  \basent{var_{\basemv{n}}}  \rbracett\).

\renewcommand{\basearg}{
\baserulehead{\basent{arg}}{::=}{\basecom{Fun arg}}\baseprodnewline
\basefirstprodline{|}{\basent{var}}{}{}{}{}\baseprodnewline
\baseprodline{|}{\basent{value}}{}{}{}{}\baseprodnewline
\baseprodline{|}{\lbracett  \basent{l_{{\mathrm{1}}}}  \texttt{=}  \basent{var_{{\mathrm{1}}}}  \texttt{;}~ \, ... \, \texttt{;}~  \basent{l_{\basemv{n}}}  \texttt{=}  \basent{var_{\basemv{n}}}  \rbracett}{}{}{}{}}

\begin{figure}
\basegrammartabular{\baseinstruction\baseinterrule
                    \baselhs\baseinterrule
                    \baserhs\baseinterrule
}
\begin{tabular}{l@{\hspace{5em}}r}
  \basegrammartabular{\basef\baseafterlastrule}&
  \basegrammartabular{\basearg\baseafterlastrule}
\end{tabular}
  \caption{\label{fig:syntax-base}Syntax of the base fragment}
\end{figure}

The type system of the base fragment is presented in figure
\ref{fig:typing-base}. In the case of instruction sequencing
\(\basent{instruction}  \texttt{;}~  \basent{instruction'}\), we do not want to restrict
\(\basent{instruction'}\) to consume exactly the values produced by
\(\basent{instruction}\). To avoid this limitation, we have added the
framing rule \(\textsc{Frame}\). This rule can be
used to extend both record types \(\basent{rty}\) and \(\basent{rty'}\) used for
typing an instruction \(\basent{instruction}\) by the same record type
\(\basent{rty''}\). This extension is performed by the join operator
\( \odot \), a partial function computing the disjoint union of two
record types.

\begin{figure}
  \begin{subfigure}{.5\linewidth}
\basedefntypingXXinstr{}
\basedefntypingXXlhs{}
\basedefntypingXXrhs{}
\basedefntypingXXf{}
\basedefntypingXXarg{}
  \end{subfigure}
  \caption{\label{fig:typing-base}Typing rules for the base fragment}
\end{figure}

\subsubsection{Operational semantics}

The semantics of the Albert base language is defined in big-step style
in Figure \ref{fig:semantics-base}. The definition of this semantic
relation is unsurprising because the base fragment is very simple and
the type system does not let much freedom at this point.

\begin{figure}
  \basedefnsJeval
  \caption{\label{fig:semantics-base}Big-step operational semantics of the base fragment}
\end{figure}

\subsection{Language extensions}

The full Albert language is obtained by adding to the base fragment
that we have just defined a series of language extensions.  The main
purpose of these extensions is to reflect all the features available
in Michelson. The only new main feature compared to Michelson is the
generalisation of the binary sum type !or! into n-ary non-recursive
variants with named constructors.

Albert's variant types generalize the !or!, !option!, and !bool! types
of Michelson. Variants are the dual of records, with the caveat that
it is not possible to construct an empty variant as Michelson does not
have an empty type it could correspond to. Variants offer two main
operations to the user: constructing a variant value using a
constructor, and pattern-matching on a variant value.

Constructors are determined by a label, and applied as a function on a
single value. When constructing a variant value, the user must
indicate the full type of the variant value because the same
constructor name is allowed to appear in different variant types. We
use the syntax \(\texttt{[}  \basent{C_{{\mathrm{1}}}}  ~\texttt{:}~  \basent{ty_{{\mathrm{1}}}}  ~\texttt{|}~ \, .. \, ~\texttt{|}~  \basent{C_{\basemv{n}}}  ~\texttt{:}~  \basent{ty_{\basemv{n}}}  \texttt{]}\) for the
variant type whose constructors are the \(\basent{C_{{\mathrm{1}}}}, \ldots, \basent{C_{\basemv{n}}}\)
where each \(\basent{C_{\basemv{k}}}\) expects an argument of type \(\basent{ty_{\basemv{k}}}\).
The types ?or a b?, ?option a?, and ?bool? in Albert are aliases for the variant types
?[Left : a | Right : b]?, ?[Some : a | None : {}]? and ?[False : {} | True : {}]? respectively.








Pattern matching can be used on variants either as a right-hand side
or as an instruction. In both cases, the Albert syntax for pattern
matching is similar to the OCaml syntax of pattern matching; for
right-hand sides, the syntax is
?match x with | pattern_1 -> rhs_1 | ... | pattern_n -> rhs_n end?.



\subsection{Example: a voting contract}
We present in figure~\ref{fig:albert-example}
a simple voting contract written in Albert. The user of the contract
can vote for a predefined set of options by sending tokens and its
choice (represented by a string) to the contract.

The storage of the contract (line 1) is a record with two fields: a
?threshold? that represents a minimum amount that must be transferred
to the contract for the vote to be considered, and an associative map,
?votes?, with strings as keys (the options of the vote) and integers
as values (the number of votes for each associated key).

If the user sends less tokens that the threshold or if the parameter
sent is not one of the options (the keys of the ?votes? map), then the
call to the contract will fail.

The contract contains two functions, ?vote? and ?guarded_vote?. Both
functions respect Michelson's call conventions: they take as input the
parameter and the storage combined and return a list of operations and
an updated storage.

?vote? checks that the parameter is one of the voting options (l.
9 and 10). If not, the contract fails (due to ?assert_some? in l.10).
Otherwise, the number of votes associated to the parameter is
increased by one (l. 11 and 12). ?vote? returns an updated
storage as well as an empty list of operations.


?guarded_vote?, the main function, checks that the amount of tokens
sent (obtained with the ?amount? primitive instruction l.21) is
greater or equal to the threshold (l.22). If so, then ?vote? is
applied. Otherwise, it fails.
\begin{figure}[htbp]
\begin{lstlisting}[language=albert,basicstyle=\footnotesize,numbers=left]
type storage_ty = { threshold : mutez; votes: map string nat }

def vote :
  { param : string ; store : storage_ty }  ->
  { operations : list operation ; store : storage_ty } =
      {votes = state; threshold = threshold } = store ;
      (state0, state1) = dup state;
      (param0, param1) = dup param;
      prevote_option = state0[param0];
      { res = prevote } = assert_some { opt = prevote_option };
      one = 1; postvote = prevote + one; postvote = Some postvote;
      final_state =  update state1 param1 postvote;
      store = {threshold = threshold; votes = final_state};
      operations = ([] : list operation)

def guarded_vote :
  { param : string ; store : storage_ty } ->
  { operations : list operation ; store : storage_ty } =
    (store0, store1) = dup store;
    threshold = store0.threshold;
    am = amount;
    ok = am >= threshold0;
    match ok with
        False f -> failwith "you are so cheap!"
      | True  t -> drop t;
          voting_parameters = { param = param ; store = store1 };
          vote voting_parameters
    end
\end{lstlisting}
\caption{\label{fig:albert-example}A voting contract, in Albert}
\end{figure}




\section{Compilation to Michelson}
\label{sec:compiler}







\subsection{Compiler architecture}

\begin{figure}
  \centering
  \includegraphics[height=4cm]{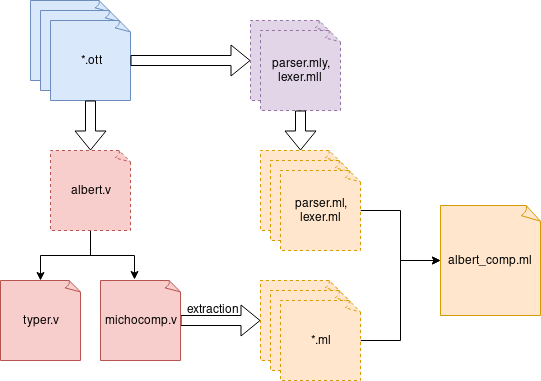}
  \caption{Compiler architecture: dashed frames designate generated
    component, solid arrows represent relevant library dependencies.}
  \label{fig:albert-compiler-arch}
\end{figure}
As we want to be able to prove the correctness of our compiler in a
near future, we decided to implement it in Coq. This allows us to easily take
advantage of Ott's definitions automatically translated to Coq, as
well as to easily compile to  Mi-Cho-Coq's AST. Moreover, using Coq's
extraction facilities,  our compiler transpiles to
OCaml code, which is more efficient and easier to use as a
library.

The global architecture of the compiler is depicted in
\ref{fig:albert-compiler-arch}. The compiler pipeline, defined using
OCaml glue code, roughly follows
a classic architecture, notwithstanding the peculiar tools used: the
lexer-parser, automatically generated from the grammar described in
Ott, produces an AST which  is then checked and annotated by the typer,
extracted from a Coq development. Then, the compilation function, also
written in Coq and extracted to OCaml, translates the typed Albert AST
into an untyped Mi-Cho-Coq AST.
Finally, the extracted Mi-Cho-Coq pretty-printer is used to produce a
string which is a Michelson program, and which the glue code dumps into
a file ready to be injected in the Tezos blockchain.

\subsubsection{Typechecker}
The type checker phase can be divided in three steps.

First, type
aliases declared by the user are replaced by their actual definition.
This will simplify the
verification of type equivalence in the next phases, as we will not
have to worry about type variables. As declared types are simple
aliases - types can't be recursively declared -- this amounts to
inlining the type aliases wherever they are found in the program.

The second step normalises type declarations by
sorting in lexicographic order both the fields of records and the
constructors of variants.

Finally, the third step checks that all defined functions are well
typed.
Currently, this type-checking proceeds in one pass from top to bottom
and it does not perform any type inference.
It checks the linearity of variable usage, the compatibility of
operands' types with their operator and the exhaustiveness of pattern
matching.
Each
instruction is  annotated with an input and output environments. The environment
being a record type, associating a type to each variable name.
One can note here that this record type is actually a
description of the Michelson stack at each point of the program where
position have been replaced by names.

The type checker is defined as a Coq function, thus is a total function.
Its implementation uses an error monad to deal with ill-typed programs.
If a program does not type check, an error message is returned
instead of the typed version of the AST.

The lack of type inference is not too much of a limitation since the
higher-level 
languages that will target Albert have enough type information to
produce the explicit type annotations that are mandatory, as for
example on variant constructors.

\subsection{Compilation scheme}

To compile an Albert program to a Michelson program, we need first
to convert Albert's types to Michelson's types and
Albert's data to Michelson's data, then to translate Albert
instructions to an equivalent Michelson sequence of instructions.

\subsubsection{Types and data}

Because Albert's primitive types reflect Michelson types, their
translation is obvious. Only the translation of records and variants
is not trivial.
Records are translated into nested pairs of values, whereas variants
are translated into a nesting of sum types.
For the sake of simplicity,  we use a comb shaped nesting,
making access to records' fields and size of variant constructor
linear in the size of the Albert type. A future task will be to provide a
syntax to control the shape of the Michelson translation or to use a
balanced tree shape.

\subsubsection{Instructions}
The compilation scheme of instructions is rather straightforward.
Projections of records fields are translated into a sequence of
projections over the relevant parts of a pair. Pattern matching over
variants are translated into a nesting of ?IF_LEFT? branchings. Each
branch of an Albert pattern-matching is translated in Michelson and
inserted in the associated position of the Michelson ?IF_LEFT?
branchings tree.

At every point of the program we memorise a mapping from variable names to
their positions in the stack.
Each operation is then translated to the equivalent operation
in Michelson, prefixed by !DIG n! operations that move the operands on
top of the stack, !n! being the index of the variables used as operands.

Function arguments are brought back on top of the stack if they are variables
and are pushed on it if they are literals.
The Michelson translation of the function is then inlined.

Assignment instructions translate into a translation of the right
hand side computation, followed by a reordering of data, guided by the
shape of the left hand side: simple variable assignments !DUG! the result
 deeper in the stack for later use, while record patterns translate to a
 pairing destruction and then some stack reorganisation.

Our mapping from variable names to stack positions is currently naive
and enforces the invariant that the elements of the stack are ordered
by the lexicographic order of the variable names. This requires too
much stack reorganisation and  will be later replaced by an optimising
placement algorithm.











\section{Related Work}
\label{sec:related-work}
Formal verification of smart contracts is a recent but active
field. The K framework has been used to formalise~\cite{kevm:2018} the
semantics of both low-level and high-level smart-contract languages
for the Ethereum and Cardano blockchains.  These formalisations have
been used to verify common smart contracts such as Casper, Uniswap,
and various implementations of the ERC20 and ERC777 standards. A
formalization of Michelson in the K framework\cite{kmichelson} is also
under development.

Note also a formalisation of the EVM in the F* dependently-typed
language~\cite{SFSAESC:POST:2018}, that was validated against the
official Ethereum test suite. This formalisation effort led to
formal definitions of security properties for smart contracts (call integrity,
atomicity, etc).

Ethereum smart contracts, written in the Solidity high-level language,
can also be certified using a translation to F*~\cite{FVSC:2016}.

The Zen Protocol~\cite{zenprotocol_whitepaper} directly uses F* as its
smart-contract language so that smart contracts of the Zen Protocol
can be proved directly in F*. Moreover, runtime tracking of resources
can be avoided since computation and storage costs are encoded in the
dependent types.

The Scilla~\cite{scilla2018} language of the Zilliqa blockchain has
been formalised in Coq as a shallow embedding. This intermediate
language is higher-level (it is based on $\lambda$-calculus) but also
less featureful (it is not Turing-complete as it does not feature
unbounded loops nor general recursion) than Michelson and Albert. Its
formalisation includes inter-contract interaction and contract
lifespan properties. This has been used to show safety properties of a
crowdfunding smart contract. Moreover, Scilla's framework for writing
static analyses~\cite{scilla2019} can be used for automated
verification of some specific properties.

In the particular case of the Tezos platform, several high-level
languages are being
developed~\cite{ligo,smartpy,fi,archetype,scaml,juvix} to ease the development of smart contracts. Formal
specification is featured in the Archetype language\cite{archetype},
the specification is then translated to the Why3 platform for
automated verification. In Juvix\cite{juvix}, dependent types can be
used to specify and verify smart contracts and resources are tracked
in a similar fashion to Albert's linear type system thanks to a
variant of quantitative type theory in Juvix's core language.



\section{Conclusion and Future Work}
\label{sec:limits-future-work}
The Albert intermediate language has been formally specified in a very
modular way using the Ott framework. This formal specification is the
unique source from which Albert's parser (written in Menhir), Albert's
typechecker and compiler (written in Coq) and the Section
\ref{sec:albert_lang} of this article (written in \LaTeX) are
generated.

The current implementation of the compiler is rather naive and we plan
to improve the performance of the produced code by sorting the values
on the Michelson stack not by the name of the corresponding Albert
variable but by their last use so that no work is performed after a
variable assignment to dive it back to its position in the stack. This
will however add some complexity in the compiler when several branches
of a pattern-matching construction are joined because we will need to
permute the stack in all but one of them to recover matching stack
types in all branches.

The Coq versions of the language specification and the compiler open
the possibility of certifying the compiler correctness and
meta-properties of the Albert language such as subject reduction and
progress. We have started proving these properties in Coq to improve
the trust in the Albert tools.

Finally, we would like to add to Albert a specification language and
support for deductive verification through the use of ghost code so
that functional verification of Tezos smart contracts can be performed
with the very high level of confidence offered by Coq and Mi-Cho-Coq
but at a higher level than Michelson.



%

\bibliographystyle{splncs04}
\bibliography{../short_references}




\end{document}